\documentclass[aps,prl,onecolumn,reprint,amsmath,amssymb,superscriptaddress, showpacs]{revtex4}
\usepackage{graphicx}
\usepackage{dcolumn}
\usepackage{bm}
\usepackage{amsmath}
\usepackage{amssymb,amsthm}
\usepackage{color}
\usepackage[sans]{dsfont}
\usepackage{algorithm}
\usepackage{algorithmic}
\usepackage{subfig}

\usepackage{epstopdf}
\usepackage{mathbbol,bbm}

\let\originalleft\left
\let\originalright\right
\renewcommand{\left}{\mathopen{}\mathclose\bgroup\originalleft}
\renewcommand{\right}{\aftergroup\egroup\originalright}

\newcommand{\ie}{\emph{i.e.}\hspace{2pt}}

\newcommand{\braces}[1]{\left\lbrace #1 \right\rbrace}

\def\r{\hat{\rho}}

\def\ham{\hat{\mathcal{H}}}
\def\partition{\mathcal{Z}}
\def\pau{\hat{\vec\sigma}}
\def\pauz{\hat{\sigma}^z}
\def\paux{\hat{\sigma}^x}

\def\fbet{F_{\mbox{\tiny Bethe}}}
\def\fkik{F_{\mbox{\tiny Kik}}}

\begin{document}
\title{Quantum Cluster Variational Method and Message Passing Algorithms Revisited}
\author{E. Dom\'{\i}nguez} 
\email{eduardo@fisica.uh.cu}
\affiliation{Group of Complex Systems and Statistical Physics, Department of Theoretical Physics, University of Havana, Cuba}

\author{Roberto Mulet} 
\email{mulet@fisica.uh.cu}
\affiliation{Group of Complex Systems and Statistical Physics, Department of Theoretical Physics, University of Havana, Cuba}

\date{\today}


\begin{abstract}
We present a general framework to study quantum disordered systems in the context of the Kikuchi's Cluster
Variational Method (CVM). The method relies in the solution of message passing-like equations for single instances 
or in the iterative solution of complex population dynamic algorithms for an average case scenario. 
We first show how a standard application of the Kikuchi's Cluster Variational Method can be easily translated to message passing equations for specific instances of the disordered system. We then present an ``ad-hoc'' extension of these equations to a 
population dynamic algorithm representing an average case scenario. At the Bethe level, these equations are equivalent to the dynamic population equations that can be derived from a proper Cavity Ansatz. However, at the plaquette approximation, the interpretation is more subtle and we discuss it  taking also into account previous results in classical disordered models. Moreover, we develop a formalism to properly deal with the average case scenario using a Replica-Symmetric ansatz within this CVM for quantum disordered systems. Finally, we present and discuss numerical solutions of the different approximations for the Quantum Transverse Ising model and the Quantum Random Field Ising model in two dimensional lattices.
\end{abstract}

\maketitle

\section{Introduction}

Exact solutions of problems involving many interacting particles in finite dimensional 
systems are very difficult to find. This is particularly true in specific complex situations where disorder is present, like in the classical Edward-Anderson model, or the quantum Anderson transition \cite{Anderson1}. When dealing with classical systems a very successful approach in the last few years has been the well known Cavity Method \cite{MP1,MP2,kabashimaBP03}, which is exact for models designed on a tree or a random graph.  Moreover, it is possible to show that the method corresponds to the Bethe approximation of the free energy for a model defined in a finite dimensional lattice\cite{yedidia,pelizzola05}, and that it is intimately connected with message passing algorithms in single instances of a specific problem\cite{Ksch,Col2,KSAT_PRE02,KSAT_JSM08}.

Although the success of the method in many models is undisputed \cite{MeMo}, it took some time to understand how to improve over this Bethe approximation for a finite dimensional disordered system \cite{tommaso_CVM,yedidia,YFW05,bolos1,bolos2,pelizzola05,kappenCVMmedical,zhouwang2012,Zhou}. The idea behind most of these improvements is based on a Cluster Variational Method\cite{Moran}, that applied to specific instances of the problem usually leads to what is called Generalized Belief Propagation (GBP) algorithm \cite{GBP_GF,dual,Ale11,Ale13}. However, it is also possible to use the same approach to the replicated free energy and then to choose the RS ansatz or the more general Parisi's hierarchical ansatz to send the number of replicas $n$ to zero \cite{tommaso_CVM,Ale13,aurell16}. Remarkably, also in this more general approximation one encounters a connection between average case predictions and the behavior of message passing algorithms in single instances of the same problem \cite{Ale13,aurell16,Ale14,Edu15}.

For quantum disordered models such a comprehension is still lacking, and even the simpler quantum models on random graphs require effort and are actively studied. A clear breakthrough in that direction was the use in \cite{Krzakala2008} of the path-integral representation of quantum spin models that was suitable to derive a closed equation where the histories of the spins constitute the proper variables to iterate. As in classical systems the method is exact in trees, however, it is much more demanding from the computational point of view. Because of its clear connection with the Cavity Method in classical systems it is usually coined {\em Quantum Cavity Method}. Somewhat similar in spirit is the approach followed in \cite{Ioffe2010a,Ioffe2010b,Dimitrova} where the quantum model is also defined in tree-like structures but is properly parametrized to simplify the solution of the corresponding closed equation. The approach can also be understood as a further approximation within the {\em Quantum Cavity Method} presented by \cite{Krzakala2008}.

Moreover, again within a Bethe approximation in \cite{Ramezanpour2012b,Biazzo2013,Biazzo2014} the authors proposed a message-passing algorithm 
to compute the Hamiltonian expectation of a quantum disordered system. The technique rests on the use of an appropriate
trial wave function and the connection of  quantum expectations to average quantities in a classical system with both local and global interactions. On the other hand, starting from a more algorithmic point of view the authors in \cite{Evans2008, Leifer2008, Poulin2008,Bilgin2009,Poulin2011,Farzad2014} proposed different versions of what they called Quantum Belief Propagation (QBP) algorithm suitable also to be used in single instances of disordered systems with tree-like topology. Here again approximations are unavoidable to guarantee a polynomial performance of the algorithm.

These approaches, however, rest on Bethe like models or Bethe like approximations to finite dimensional systems and 
are difficult to extend to more general scenarios. Fortunately, already Morita and Tanaka \cite{Morita57a,Morita57b,Tanaka94,Tanaka94b,Morita94b} 
developed a general approach to derive a set of self-consistent equations within a Cluster Variational Method formalism. However, as far as we know these equations were mainly studied in homogeneous systems and only formally presented for disordered models.

In this work we first take profit of this derivation and the parametrization done in \cite{Dimitrova} to study single instance implementations of message passing algorithms for quantum disordered systems at both the Bethe level and the plaquette approximation. Moreover, inspired by  previous results on classical models \cite{tommaso_CVM,Ale13,Edu15}, we propose a set of fixed point equations to describe these quantum models without the specification of the instance and compare both approaches. At the Bethe level these equations are a generalization of the dynamic population equations already derived in \cite{Morita57a,Morita57b}. At the plaquette level, to our knowledge, this is the first time that these equations are presented for quantum models. Finally we present how to properly generalize the cluster variational method for quantum disordered models within the RS-ansatz in an average case scenario. These approaches were tested in the Quantum Ising model in a transverse field and in the Quantum Ising model with a random field.

The rest of the work is organized in the following way. In the next section we present the models used to test the different approximations. This will make clear the kind of models of interest and help to simplify the notation below. Then we will present the Cluster Variational Method in the form derived by Morita and Tanaka \cite{Morita57a,Morita57b,Tanaka94,Tanaka94b,Morita94b} in their seminal works but using a modern notation and connecting it to message passing equations. We do this for the Bethe and the plaquette approximation. Then we show how to derive similar equations but taking into account the average over the disorder. Finally we present and discuss the numerical results of the different approximations for the two models under study.

\section{Models}

The Quantum Ising Model in a transverse field is one of the most basic models displaying quantum phase transitions. Under this name we identify  a general class of models described by the Hamiltonian:

\begin{equation}
 \ham = - \sum_{(ij)} J_{ij} \paux_i \paux_j - \sum_{(i)} h_i \pauz_i
 \label{eq:hamiltonian}
\end{equation}

\noindent where the choice of the interacting pairs $(i,j)$ defines the underlying lattice. The properties of this system obviously
depend on the topological structure as well as on the distribution of interaction constants $J_{ij}$ and local fields $h_i$. 
Popular choices for the fields and interactions include the ferromagnetic case, where $J_{ij}=J$ is uniform and positive everywhere and the
local field is also uniform. More interesting is the so called Random Field Ising Model (RFIM) where interactions are homogeneous too but local fields fluctuate from site to site according to a given distribution $P_h(h_i)$.


In contrast to the classical version, the local fields applied perpendicularly to the easy direction of the material included in Eq. \eqref{eq:hamiltonian} result in non-commuting terms that make the Hamiltonian difficult to diagonalize. The basis with a definite $x$ direction of each spin is no longer an eigenstate of the  system, meaning that transverse fields introduce quantum fluctuations that can destroy any long range order for sufficiently strong field intensities.

An exact solution for this kind of systems has been found only for special cases, mainly mean field and fully connected models or
very particular tree-like topologies such as a random lattice\cite{Krzakala2008}.
For the latter, the Quantum Cavity Method is the most successful technique, but solutions for finite dimensional lattices remain elusive, and this Quantum Cavity Method represents just an approximation at the Bethe level of the actual problem\cite{Ioffe2010a,Ioffe2010b,Dimitrova}.
In finite dimensional lattices a quantum extension of the CVM is suitable because of the geometric
and periodic character of the structure. We  follow this line of thought in the next sections to obtain a description of both, the ferromagnetic and RFIM at the level of the Bethe (pairwise) and Kikuchi (plaquette) free energy approximations for a bi-dimensional square lattice. Extensions to other lattice models should follow straightforwardly.


\section{Quantum Cluster Variational Method: Single Instances and Message Passing}

The  Cluster Variational Method \cite{kikuchi,pelizzola05,yedidia,Morita57a,Morita57b,Tanaka94,Tanaka94b,Morita94b,Moran} relies on a constrained minimization procedure of a region-based
free energy functional. As a result, it is possible to get estimates for the local probability distributions of the system
under study; distributions that are otherwise hard to obtain because of the well known difficulty of tracing the complete
distribution over the exponential number of allowable states in the large $N$ limit. 

For quantum problems the equivalent of local probability distributions are the projections of the full system density matrix
on the subset of variables of each region. With these local density operators it is possible to write a region free energy
and repeat essentially the same classical minimization. 

\subsection{Bethe approximation}

In the specific case of pairwise interactions at the Bethe level,  
the variational parameters of the problem are the pair and site density operators, denoted $\r_l^{(ij)}$ and $\r_s^{(i)}$ respectively.
The Bethe free energy $\fbet$ is thus written in terms of these operators as a sum of weighted contributions of all pairs and sites
of the interaction network or lattice:
\begin{equation}
\fbet = \sum_{(ij)} c_l F_l^{(ij)} +  \sum_{(i)}^N c_s F_s^{(i)}
 \label{eq:free_energy_Bethe}
\end{equation}
where 
\begin{equation}
 F_l^{(ij)} = Tr[\ham_l^{(ij)} \r_l^{(ij)}] + \dfrac{1}{\beta} Tr[ \r_l^{(ij)} \ln  \r_l^{(ij)}]
 \label{eq:free_energy_link}
\end{equation}
can be regarded as the free energy of the pair $l\equiv(ij)$ and the other term represents the contribution of the individual sites:

\begin{eqnarray}
\label{eq:free_energy_spin}
 F_s^{(i)} &=& Tr[\ham_s^{(i)} \r_s^{(i)}] + \dfrac{1}{\beta} Tr[ \r_s^{(i)} \ln  \r_s^{(i)}]
\end{eqnarray}

The prefactors $c_l$ and $c_s$ take integer values such that the contribution of each variable to the total free energy is counted only
once. These are the two first terms in a cumulant expansion of the total free energy \cite{Tanaka94}.
For the 2D square lattice $c_l = 1$ and $c_s = -3$. 
The minimization of $\fbet$ is performed under a set of constraints enforcing normalization and consistency of the set of operators
$\{\r_l^{(ij)}\}$ and $\{\r_s^{(i)}\}$:

\begin{eqnarray}
 \nonumber
 Tr[\r_s^{(i)}] &=& 1 \;\; \forall i\\
 \label{eq:consistency_equation} 
 Tr[\r_l^{(ij)}] &=& 1  \;\; \forall \;\; l=(ij)\\
 \nonumber
 \r_s^{(i)}&=&Tr_j[\r_l^{(ij)}]\;\;\; \forall l, \;\forall i \in l 
\end{eqnarray}

We can use now Lagrange multipliers to put together the objective function $\fbet$ and the restrictions \eqref{eq:consistency_equation}. Up to this point there is no particular relevance in the choice of the basis for the 
Hilbert state space. In fact, the Lagrange function for this problem can be nicely written in a basis-independent way:
\begin{equation}
 \mathcal{L}[\{\r_l^{(ij)}\},\{\r_s^{(i)}\}] = \fbet + \sum_{(i)}^N \alpha_i (Tr[\r_s^{(i)}] - 1)
 +   \sum_{(ij)} \alpha_l (Tr[\r_l^{(ij)}] - 1)
 \nonumber
 +   \sum_{(i)}^N \sum_{l \in \mathcal{P}(i)} Tr \left[\hat{\lambda}_{l\rightarrow i}\left(\r_s^{(i)} - Tr_j[\r_l^{(ij)}]\right)\right]
 \label{eq:lagrange_function}
\end{equation}

In this last equation \eqref{eq:lagrange_function} the different $\alpha$'s are real numbers and $\hat{\lambda}_{l\rightarrow i}$
are Hermitian operators acting on the single site Hilbert spaces. The stationarity condition for $\mathcal{L}$ is obtained by setting to zero the linear part in $\epsilon$ of the increment $$\delta \mathcal{L}\equiv \mathcal{L}(\left\lbrace \hat{\rho} + \epsilon \;\delta \hat{\rho}\right\rbrace ) - \mathcal{L}(\left\lbrace \hat{\rho}\right\rbrace ).$$

The resulting equations are the operator version of the belief propagation (BP) \cite{yedidia} equations for the local distributions in terms of the Lagrange multipliers:

\begin{eqnarray}
 \r_s^{(i)} &=& \dfrac{1}{\mathcal{Z}_s^{(i)}} \exp -\beta\left(\ham_s^{(i)} - \dfrac{1}{c_s - 1} \sum_{l \in \mathcal{P}(i)}  \hat{\lambda}_{l\rightarrow i}\right)\\
 \r_l^{(ij)} &=& \dfrac{1}{\mathcal{Z}_l^{(ij)}} \exp -\beta\left(\ham_l^{(ij)} - \hat{\lambda}_{l\rightarrow i} - \hat{\lambda}_{l\rightarrow j}\right)
\end{eqnarray}

It is customary to make the linear transformation: 
\begin{equation}
 \hat{\lambda}_{l\rightarrow i} =  \sum_{l' \in \mathcal{P}(i)\setminus l} \hat{u}_{l'\rightarrow i}
\end{equation}
where the sum includes all links containing $i$, denoted as $\mathcal{P}(i)$, except the link $l$ itself.
This substitution gives the more familiar BP-like structure:

\begin{eqnarray}
   \label{eq:quantum_beliefs_spin}
   \r_s^{(i)} &=& \dfrac{1}{\mathcal{Z}_s^{(i)}} \exp -\beta\left(\ham_s^{(i)} - \sum_{l' \in \mathcal{P}(i)}  \hat{u}_{l'\rightarrow i}\right)\\
   \label{eq:quantum_beliefs_link}
   \r_l^{(ij)} &=& \dfrac{1}{\mathcal{Z}_l^{(ij)}} \exp -\beta\left(\ham_l^{(ij)} - \sum_{l' \in \mathcal{P}(i) \setminus l} \hat{u}_{l'\rightarrow i} - \sum_{l'' \in \mathcal{P}(j) \setminus l} \hat{u}_{l'' \rightarrow j}\right)
\end{eqnarray}

The operator $\hat{u}_{l\rightarrow i}$ represents the effective interaction of the spin $i$ with its neighbor $j$, both forming the link $l=(ij)$. Since the site-site interaction in \eqref{eq:hamiltonian} is directed
along the  $x$ direction we chose to parametrize this operator as $\hat{u}_{l\rightarrow i} \equiv u_{l\rightarrow i} \paux_i$.
This form resembles the projected cavity solution in \cite{Dimitrova}. In this work the authors make a recursive construction
in a tree  to obtain a self-consistent equation for a cavity field. In our parametrization, $u_{l\rightarrow i}$ (without a hat) can be interpreted again as a kind of cavity magnetic field, in the sense that it stands for the interaction of spin $i$ with the portion of the network growing in the direction of link $l$. 

The set of $\left\lbrace u_{l\rightarrow i}\right\rbrace $ fields can be determined by a fixed point iteration after plugging \eqref{eq:quantum_beliefs_spin} and  \eqref{eq:quantum_beliefs_link} into the consistency conditions in \eqref{eq:consistency_equation}. The procedure is analogous to the use of the BP algorithm in the classical case. This time though, we have a set of coupled operator equations:

\begin{eqnarray}
   \label{eq:consistency_equation_bethe}
   \dfrac{1}{\mathcal{Z}_s^{(i)}} 
   \exp -\beta\left(\ham_s^{(i)} - \paux_i \sum_{l' \in \mathcal{P}(i)}  u_{l'\rightarrow i}\right) &=&
   \dfrac{1}{\mathcal{Z}_l^{(ij)}} 
   Tr_j[\exp -\beta\left(\ham_l^{(ij)} -  \paux_i  \sum_{l' \in \mathcal{P}(i) \setminus l} u_{l'\rightarrow i} 
   -  \paux_j \sum_{l'' \in \mathcal{P}(j) \setminus l} u_{l'' \rightarrow j}\right)]
\end{eqnarray}
\noindent where the (local) partition functions $\mathcal{Z_R}$ are fixed from the normalization condition $Tr[\r_R] = 1$. Spin $i$, for example, will have the following normalization:
\begin{equation}
 \mathcal{Z}_s^{(i)} = Tr[ \exp -\beta\left(\ham_s^{(i)} - \paux_i \sum_{l' \in \mathcal{P}(i)}  u_{l'\rightarrow i}\right) ]
 \label{eq:partition_function_spin_single_instance}
\end{equation}

Then, instead of working in \eqref{eq:consistency_equation_bethe} directly with operators that are hard to translate into actual numerical values it is convenient to write the consistency between them by matching their moments. This is, we make use of the relation $m^x_i \equiv Tr[\paux_i \r_s^{(i)}] = Tr[\paux_i \r_l^{(ij)}] $ a relation that is sufficient for our purposes of finding $u_{l\rightarrow i}$. 

\begin{figure}[h]
 \centering
 \includegraphics[width=0.5\textwidth,keepaspectratio=true]{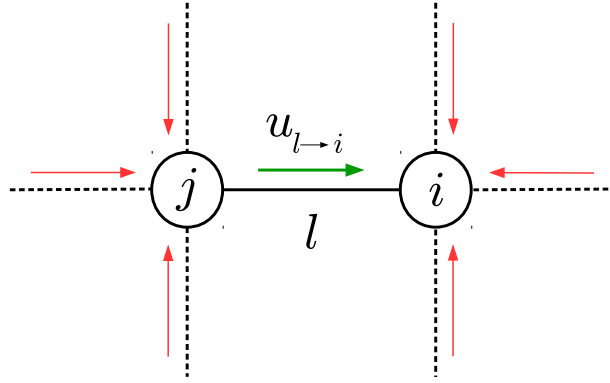}
 \caption{Message passing in the Bethe approximation. To calculate the field $u_{l\rightarrow i}$ (green thick line) at time $t+1$, we need  to sample all the fields acting on the link (red thin lines) at the previous iteration step, $t$. Using the same external
 messages in red we can also determine $u_{l\rightarrow j}$ (not shown in the figure).}
 \label{fig:message_passing_bethe}
\end{figure}

The algorithmic procedure to solve for all the values in a given lattice is the following. First, select at random
a link region $l\equiv(ij)$. Then, using all the $u_{l'\rightarrow i}$ and $u_{l''\rightarrow j}$ acting on each one of the spins from
\textit{outside} $l$, form the local density operator $\r_l^{(ij)}$ and find the magnetization of spin $i$, $m^x_i$. Now focus on
the expression for $\r_s^{(i)}$. All cavity fields in the exponent were used in the link equation except precisely $u_{l\rightarrow i}$, the cavity field of link $l$ on spin $i$. Its value is now found numerically as the root of the equation  $m^x_i - Tr[\paux_i \r_s^{(i)}]=0$. This step is repeated many times on different links and spins pairs until the value of $u_{l\rightarrow i}$ changes
less than a certain tolerance everywhere in the lattice. The process is schematically explained in {\bf Fig.\ref{fig:message_passing_bethe}}. In 
thinner red lines appear the messages taken at step $t$ to find a new value (green, thick line) at step $t+1$. The reader familiar
with the BP algorithm may notice that in this case we need to keep the cavity fields on $i$ during the calculation, whereas
in the classical case they can be canceled out in both sides of \eqref{eq:consistency_equation_bethe} due to the commutation
properties of the effective Hamiltonian.

\subsection{Kikuchi approximation}

For finite dimensional lattices it is important to take into account explicitly the existence of short loops, which are completely disregarded by the Bethe approximation \cite{yedidia}. A sound improvement of the Bethe choice of regions could be obtained by including larger regions into the free energy. The simplest generalization for a square lattice is precisely the inclusion of plaquette regions, each formed by the four spins of the elementary cell. The Kikuchi free energy $\fkik$ will have an extra term with respect to \eqref{eq:free_energy_Bethe}, comprising the contribution of all plaquettes: 

\begin{equation}
 \fkik = \sum_{(ijkm)} c_p F_p^{(ijkm)} + \sum_{(ij)} c_l F_l^{(ij)} +  \sum_{(i)} c_s F_s^{(i)}
 \label{eq:free_energy_Kikuchi}
\end{equation}
The free energy of a plaquette $(ijkm)$ is defined similarly to \eqref{eq:free_energy_link} and \eqref{eq:free_energy_spin} by means of
a plaquette density operator $\r_p^{(ijkm)}$. The prefactors take the values $c_p=1$, $c_l=-1$ and $c_s =1$ in this case. Constrained minimization of $\fkik$ is technically similar to the Bethe case except for the
use of some extra Lagrange multipliers $\hat{U}_{p'\rightarrow l}$ that enforce marginalization of plaquettes over link distributions.
These new multipliers are Hermitian operators acting on  the two-spins Hilbert space corresponding to the spins in each link. They represent an effective $x$-directed interaction of the spins in the link with the rest of the lattice. Hence, we write them in the following way:

\begin{equation}
 \hat{U}_{p\rightarrow l} = U_{p\rightarrow l} \paux_i \paux_j +   u_{p\rightarrow i} \paux_i +   u_{p\rightarrow j} \paux_j
\end{equation}

This is not the most general expression for an operator in the product space of two spins. This choice is based on the fact that the spin to spin interaction lies always in the OX direction.
The value of the Lagrange parameters should be obtained from the marginalization conditions \eqref{eq:consistency_equation} and the equivalent for plaquettes. For the local density operators of each region we get:


\begin{eqnarray}
\label{eq:belief_U1}
\r_s^{(i)} &=& \dfrac{1}{\mathcal{Z}_s^{(i)}} \exp -\beta\left(\ham_s^{(i)} -  \sum_{l' \in \mathcal{P}(i)}  \hat{u}_{l'\rightarrow i}\right)\\
\label{eq:belief_U2}
  \r_l^{(ij)} &=&\dfrac{1}{\mathcal{Z}_l^{(ij)}}  \exp -\beta\left(\ham_l^{(ij)} 
		- \sum_{\substack{i'\in \mathcal{D}(l)\\ l'\in \mathcal{P}(i')\setminus l}} \hat{u}_{l'\rightarrow i'} 
		- \sum_{p\in \mathcal{P}(l)} \hat{U}_{p\rightarrow l}\right)\\
\label{eq:belief_U3}
  \r_p^{(ijkm)} &=& \dfrac{1}{\mathcal{Z}_p^{(ijkm)}} \exp -\beta\left(\ham_p^{(ijkl)}
		  - \sum_{\substack{i\in \mathcal{D}(p) \\ l'\in \mathcal{P}(i)\setminus \mathcal{D}(p)}} \hat{u}_{l'\rightarrow i} 
		  - \sum_{\substack{l\in \mathcal{D}(p) \\ p'\in \mathcal{P}(l)\setminus p }} \hat{U}_{p'\rightarrow l} 
		  \right)
\end{eqnarray} 
where $p,l,i$ are indexes corresponding to plaquette, link and site regions respectively. For a region $R$ the set $\mathcal{D}(R)$ contains
all its sub-regions and the set $\mathcal{P}(R)$ is populated with all the regions to which $R$ belongs. For example, for the link $l\equiv (i,j)$, $\mathcal{D}(l)$ contains the two spin regions $i$ and $j$, whereas $\mathcal{P}(l)$ includes all the plaquettes intersecting on $l$.

The fixed point iterations are again performed via moment matching. This is, the $m_x$ magnetization predicted by $\r_s^{(i)}$ must be consistent with $\r_l^{(ij)}$ and this last operator must also produce the same magnetization and correlation as the parent plaquettes $\r_p^{(ijkm)}$:

\begin{eqnarray}
 Tr[\paux_i \r_s^{(i)}] &=& Tr[\paux_i \r_l^{(ij)}] \label{eq:mi_consistency_spin}   \\
 Tr[\paux_i \r_l^{(ij)}] &=& Tr[\paux_i \r_p^{(ijkm)}]  \label{eq:mi_consistency} \\
 Tr[\paux_j \r_l^{(ij)}] &=& Tr[\paux_j \r_p^{(ijkm)}]  \label{eq:mj_consistency} \\
 Tr[\paux_i \paux_j \r_l^{(ij)}] &=& Tr[\paux_i \paux_j \r_p^{(ijkm)}]  \label{eq:corr_consistency}  
 \end{eqnarray}

The set \eqref{eq:mi_consistency_spin}-\eqref{eq:corr_consistency} gives the solution for the cavity fields implicitly. For future use it is convenient to formally define the functions that would give this fields in a  explicit way:
\begin{eqnarray}
 \nonumber
 u_{l\rightarrow i} &=& \hat  u_{l\rightarrow i}(\#)\\
  \label{eq:explicit_update}
 u_{p\rightarrow i} &=& \hat  u_{p\rightarrow i}(\#)\\ 
 \nonumber
 U_{p\rightarrow l} &=& \hat  U_{p\rightarrow l}(\#)
\end{eqnarray}

\begin{figure}[h]
 \centering
 \includegraphics[width=0.5\textwidth,keepaspectratio=true]{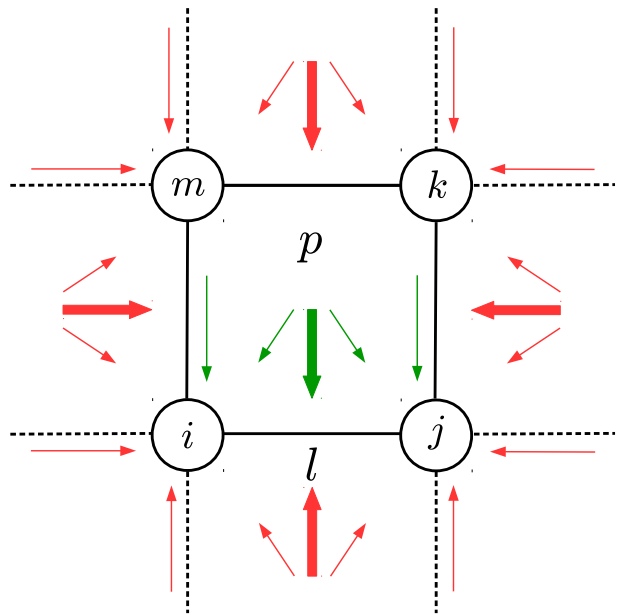}
 \caption{Message passing in the Kikuchi approximation. Sampling the fields external to plaquette $p$ (red arrows) at step $t$ we can determine the messages inside it (in green) at time $t+1$.
 In the figure, arrows starting in the center of a plaquette represent the triad $(U_{p\rightarrow l},u_{p\rightarrow i},u_{p\rightarrow j})$, the thick one corresponding to the correlation field $U_{p \rightarrow l}$. Other arrows, parallel to plaquette edges, are link-to-spin messages, $u_{l\rightarrow i}$.}
 \label{fig:message_passing_kikuchi}
\end{figure}

The algorithmic details for the implementation of the fixed point iteration are very similar to the Bethe case. We take a random plaquette with the corresponding external cavity fields at step $t$ (see Fig. (\ref{fig:message_passing_kikuchi})) and evaluate the RHS of \eqref{eq:mi_consistency} and \eqref{eq:corr_consistency} for every connected pair and single spin in the plaquette. Then, by a fixed point iteration, we make
the LHS of the same equations be consistent with the plaquette prediction. The consistency equation \eqref{eq:mi_consistency_spin}
is used too at this step. This way we find the internal fields of the plaquette at step $t + 1$ in terms of the external ones at $t$.

Once the approximated marginal distributions $\r$ are known, the local contributions to the free energy can easily be obtained through the 
normalization factors of \eqref{eq:belief_U1}-\eqref{eq:belief_U3}:

\begin{equation}
 -\beta \fkik =  \left[ \sum_{(ijkm)} c_p \ln \mathcal{Z}_p^{(ijkm)} + \sum_{(ij)} c_l \ln\mathcal{Z}_l^{(ij)} +  \sum_{(i)} c_s\ln\mathcal{Z}_s^{(i)} \right]
 \label{eq:free_energy_Kikuchi_partition_single_instance}
\end{equation}

 Finally let us discuss the connection of the formalism to the classical GBP. The classical results are obtained effortlessly from \eqref{eq:belief_U1}-\eqref{eq:corr_consistency} when the transverse field is zero. In these cases, all terms in the exponentials conmute and the cavity fields that appear in both sides  of the consistency equations can be cancelled out. When conmutation is important though, the terms do not cancel and most be tacking into account.

\section{Quantum Cluster Variational Method: Average case scenario}

In the  previous sections we discussed how to deal with single instances of disordered systems withing a Quantum Cluster Variational method. 
More frequently in physics one expects to be able to average over the disorder right from the beginning. Inspired by the replica-CVM methodology introduced in \cite{tommaso_CVM} for the classical version we now try to perform the average case calculations for disordered Ising quantum models.


The replica trick is a general framework that in principle allows to account for different degrees of complexity in
the structure of the state space. Here we present the solution of the problem within the Replica Symmetric (RS) approximation that assumes
that a single state dominates the thermodynamics of the problem. Within this approximation we studied the RFIM at the level of the Bethe and Kikuchi approximations starting from  \eqref{eq:hamiltonian}  with $h_i$ being i.i.d random variables in the interval $[0,h)$. The extension to more general ansatz follows directly from the work \cite{tommaso_CVM} and the approach presented here.



As a starting point let us consider an alternative approach to find the CVM free energy of a single instance. With a given realization of the disorder $\left\lbrace h \right\rbrace$ one way to find the region free energy of the system
is to minimize a variational expression that is equivalent to the Lagrange function discussed in previous sections:

\begin{eqnarray}
 \label{eq:free_cvm_variational}
 F_{\text{CVM}}^{\text{var}}(\left\lbrace h\right\rbrace,\left\lbrace u\right\rbrace) &=&  -\dfrac{1}{\beta} \sum_R c_R \ln\mathcal{Z}_R(\left\lbrace u\right\rbrace_R)\\
 F_{\text{CVM}}(\left\lbrace h\right\rbrace) &=& \min_{\left\lbrace u\right\rbrace}\left[ F_{\text{CVM}}^{\text{var}}(\left\lbrace h\right\rbrace,\left\lbrace u\right\rbrace) \right]
\end{eqnarray}
Here $\left\lbrace u\right\rbrace_R$ is the subset of cavity fields appearing in the expressions for $\mathcal{Z}_R$, the normalization
constant of each region distribution. The minimization process makes this set of fields depend in principle on all the external parameters
$\left\lbrace h\right\rbrace$:
\begin{equation}
 F_{\text{CVM}}(\left\lbrace h\right\rbrace) = -\dfrac{1}{\beta} \sum_R c_R \ln\mathcal{Z}_R(\left\lbrace u\right\rbrace_{R,\left\lbrace h\right\rbrace})
 \label{eq:free_cvm_minimized}
\end{equation}

The average free energy density is now defined in the thermodynamic limit as the free energy per spin after averaging over the distribution of
$\left\lbrace h\right\rbrace$:
\begin{equation}
 \langle f_{\text{CVM}} \rangle \equiv \lim_{N\rightarrow \infty} \dfrac{1}{N} \langle F_{\text{CVM}}(\left\lbrace h\right\rbrace) \rangle_{\left\lbrace h\right\rbrace}
 \label{eq:average_free_cvm}
\end{equation}
Here $F_{\text{CVM}}$  stands for the Bethe or Kikuchi approximations $\fbet$ and $\fkik$ defined in \eqref{eq:free_energy_Bethe} and \eqref{eq:free_energy_Kikuchi} or any other valid region based free energy approximation. Putting \eqref{eq:free_cvm_minimized} into \eqref{eq:average_free_cvm}
we see that the problem now reduces to averaging the logarithm of the local partition functions:

\begin{equation}
 \langle f_{\text{CVM}} \rangle= \lim_{N\rightarrow \infty} 
 -\dfrac{1}{\beta N}    
 \sum_R c_R
 \langle 
 \ln\mathcal{Z}_R(\left\lbrace u\right\rbrace_{R,\left\lbrace h\right\rbrace})
 \rangle_{\left\lbrace h\right\rbrace}
 \label{eq:average_free_cvm_1}
\end{equation}

This is not an easy task, mainly because we do not know the analytic dependence of the cavity fields on the realization of the disorder;
remember that the cavity fields are found via a fixed point iteration. To overcome this problem let us take a step back, return to the
variational character of the expression \eqref{eq:free_cvm_variational} and define:

\begin{equation}
 \langle f_{\text{CVM}} \rangle^{\text{var}}= \lim_{N\rightarrow \infty} 
 -\dfrac{1}{\beta N}    
 \sum_R c_R
 \langle 
 \ln\mathcal{Z}_R(\left\lbrace u\right\rbrace_{R})
 \rangle_{\left\lbrace h\right\rbrace_R}
 \label{eq:average_free_cvm_2}
\end{equation}
The difference between \eqref{eq:average_free_cvm_1} and \eqref{eq:average_free_cvm_2} is that in the latter the fields are free parameters
to be optimized. Notice also that each region depends only on the external fields acting locally.
Since the cavity fields are independent parameters (to be fixed later by a minimization process) the average over all the set $\left\lbrace h\right\rbrace$ reduces to only the fields in the region $\left\lbrace h\right\rbrace_R$. Another useful manipulation is to split the sum over $R$ according to the kind of region $\sum_R[\cdot] = \sum_{r\in\left\lbrace s,l,p \right\rbrace} \sum_{R_r} [\cdot] $. Here $r$ is an index that 
goes over the types of regions used in the approximation and $R_r$ labels different regions of the same kind. Also, let us define $b_r N$ as the number of regions of type $r$ in the system. This way we arrive to an expression that has the form of an average over the disorder and the cavity fields:
\begin{eqnarray}
 \langle f_{\text{CVM}} \rangle^{\text{var}} &=&  -\dfrac{1}{\beta} \sum_{r\in\left\lbrace s,l,p \right\rbrace} 
 b_r c_r \lim_{N\rightarrow \infty}  \langle \dfrac{ \sum_{R_r} \ln\mathcal{Z}_{R_r}(\left\lbrace u\right\rbrace_{R_r}) }{b_r N}\rangle_{\left\lbrace h\right\rbrace_{R_r}}\\
 &=&  -\dfrac{1}{\beta} \sum_{r\in\left\lbrace s,l,p \right\rbrace} 
 b_r c_r \langle  \ln\mathcal{Z}_{r}(\left\lbrace u\right\rbrace_{r}) \rangle_{\left\lbrace h\right\rbrace_{r},\left\lbrace u\right\rbrace_{r}}
 \label{eq:average_free_cvm_3}
\end{eqnarray}

Notice that the sum in \eqref{eq:average_free_cvm_3} is not extensive anymore; for the Bethe and Kikuchi approximation it has 2 and 3 terms respectively. This is now a functional on the cavity field distribution and at the end of the calculations it should be minimized.
For each region type the set $\left\lbrace u\right\rbrace_{r}$ includes a different number of link-to-spin and plaquette-to-link fields; in order to
make this explicit it is convenient to use the notation $\left\lbrace u\right\rbrace_{r}\equiv  \vec u_l,\vec u_p, \vec U_p$.

The RS anzats implies that the cavity fields in the expression for $\mathcal{Z}_{r}$ are described by a certain distribution $P_r(\vec u_l,\vec u_p, \vec U_p)$: 

\begin{equation}
 \langle \ln\mathcal{Z}_{r}\rangle_{\left\lbrace h\right\rbrace_{r},\left\lbrace u\right\rbrace_{r}} = \int d\vec h d\vec u_l d\vec u_p d\vec U_p
 P_h(\vec h) P_r(\vec u_l,\vec u_p, \vec U_p) \ln\mathcal{Z}_{r}(\vec h, \vec u_l,\vec u_p, \vec U_p)
 \label{eq:average_partition_function}
 \end{equation}

This is the same expression that is obtained by the explicit replica symmetric calculations in \cite{tommaso_CVM}. Symmetry breaking considerations may define an average \eqref{eq:average_partition_function} that include distributions of distributions in the case of 1RSB or more levels in general.

A further assumption in our RS calculation is that the joint $P_r(\vec u_l,\vec u_p, \vec U_p)$ is factored out in terms of simpler distributions. In general we consider that each $u_{l\rightarrow i}$ in the terms $u_{l\rightarrow i} \paux_i$ is independently distributed with probability $q(u_{l\rightarrow i})$ and the same for the fields in $\hat{U}_{p\rightarrow l} = U_{p\rightarrow l} \paux_i \paux_j +   u_{p\rightarrow i} \paux_i +   u_{p\rightarrow j} \paux_j$, distributed according to $Q(U_{p\rightarrow l},u_{p\rightarrow i}, u_{p\rightarrow j})$. This simplifications are justified for Bethe lattices but for finite dimensional problems remain an approximation. The disordered fields $h_i$ are distributed independently so their distribution
factorizes too.

Our task now is to find the $q(u)$ and $Q(U,u_1,u_2)$ such that the functional $\langle f_{\text{CVM}} \rangle^{\text{var}}$ reaches a minimum.
We have to solve the stationarity conditions:
\begin{eqnarray}
 \dfrac{\partial \langle f_{\text{CVM}} \rangle^{\text{var}} }{\partial q} &=& 0\\
 \dfrac{\partial \langle f_{\text{CVM}} \rangle^{\text{var}} }{\partial Q(U,u_1,u_2)} &=& 0
 \label{eq:stationarity_condition_kikuchi}
\end{eqnarray}

\subsection{Bethe approximation}

Let us start by writing down with detail the Bethe average case calculation. The (variational) average free energy for a network 
with fixed connectivity $c$ is, using \eqref{eq:average_free_cvm_3}:

\begin{equation}
 -\beta f_{\text{Bethe}} = \dfrac{c}{2} \langle \ln \partition_l \rangle - (c-1) \langle \ln \partition_s \rangle
 \label{eq:variational_free_energy_bethe}
\end{equation}

The averages in \eqref{eq:variational_free_energy_bethe} are over the external field and the cavity fields. The explicit expressions for each
term are:

\begin{equation}
 \langle \ln \partition_l \rangle = \int dh_i dh_j \left[\prod_{k}^{c-1} du_{k\rightarrow i}  \prod_{m}^{c-1} du_{m\rightarrow j}\right]
 P_h(h_i) P_h(h_j) P_u^l(\braces{u_{k\rightarrow i}},\braces{u_{m\rightarrow i}})  \ln \partition_l (\braces{u_{k\rightarrow i}},\braces{u_{m\rightarrow j}},h_i,h_j)
 \label{eq:average_partition_link_bethe}
\end{equation}
and
\begin{equation}
 \langle \ln \partition_s \rangle = \int dh_i \left[\prod_{k}^{c} du_{k\rightarrow i}\right]
 P_h(h_i) P_u^s(\braces{u_{k\rightarrow i}})  \ln \partition_s (\braces{u_{k\rightarrow i}},h_i)
 \label{eq:average_partition_spin_bethe}
\end{equation}

The partition function for the link and spin regions are as usual defined as the trace of the Boltzmann factor
for the corresponding effective hamiltonian:

\begin{eqnarray}
 \partition_s (\braces{u_{k\rightarrow i}},h_i) &=&Tr\left[ \exp -\beta\left(\ham_s^{\text{eff}}(\pau_i) \right) \right]\\
 \ham_s^{\text{eff}}(\pau_i)&=& \ham_s(\pauz_i) - \paux_{i} \sum_{l'\in\mathcal{P}(i)}   u_{l'\rightarrow i}\\
 \partition_l (\braces{u_{k\rightarrow i}},\braces{u_{m\rightarrow j}},h_i,h_j) &=& Tr\left[\exp -\beta
      \left(       
      \ham_l^{\text{eff}}(\pau_{i},\pau_{j})
      \right)\right] \\
\ham_l^{\text{eff}}(\pau_{i},\pau_{j}) &=& 
       \ham_l(\pau_{i},\pau_{j}) 
      - \paux_{i} \sum_{k \in \mathcal{P}(i)\setminus l} u_{k\rightarrow i}
      - \paux_{j} \sum_{m \in \mathcal{P}(j)\setminus l} u_{m\rightarrow j}
\end{eqnarray}

The important quantity in \eqref{eq:average_partition_link_bethe} and \eqref{eq:average_partition_spin_bethe} are the distributions $P_u^s$ and $P_u^l$. In the classical version of this 
calculation one assumes that cavity fields are uncorrelated and these distributions factorize. This is literally true
for random networks and only an approximation for lattices with short loops. The factorized forms we consider are:
\begin{equation}
 P_u^s(\braces{u_{k\rightarrow i}}) = \prod_{k}^{c} q(u_{k\rightarrow i})
 \label{eq:factorization_prob_spin}
\end{equation}
and 
\begin{equation}
 P_u^l(\braces{u_{k\rightarrow i}},\braces{u_{m\rightarrow i}})  = \prod_{k}^{c-1} q(u_{k\rightarrow i}) \prod_{m}^{c-1} q(u_{m\rightarrow j})
\end{equation}

The next step is to plug everything into the free energy and use the stationarity condition $\frac{\partial  f_{\text{Bethe}} }{\partial q(u_{k_0\rightarrow i})} = 0 $. A simple functional derivative shows that the minimization implies that:

\begin{equation}
 \langle \ln \partition_s(u_{k_0\rightarrow i}) \rangle = \langle \ln \partition_l(u_{k_0\rightarrow i}) \rangle
 \label{eq:minimization_free_1}
\end{equation}
where the explicit dependence on $u_{k_0\rightarrow i}$ means that this variable is not averaged out. Differentiating both sides we get a 
nice relation between the average magnetization predicted by the link and the spin terms:

\begin{equation}
 \langle m_s(u_{k_0\rightarrow i}) \rangle = \langle m_l(u_{k_0\rightarrow i}) \rangle
 \label{eq:minimization_free_2}
\end{equation}

Multiplying both sides by $q(u_{k_0\rightarrow i})$ and integrating to average out also $u_{k_0\rightarrow i}$:

\begin{equation}
 \langle m_s \rangle = \langle m_l\rangle
 \label{eq:minimization_free_3}
\end{equation}
where averages on left and right hand sides are done using expressions similar to \eqref{eq:average_partition_spin_bethe} and 
\eqref{eq:average_partition_link_bethe} respectively:

\begin{equation}
 \langle m_s \rangle = \int dP_h(h_i) \left[\prod_{k}^{c} dq(u_{k\rightarrow i})\right]
 m_s (\braces{u_{k\rightarrow i}},h_i)
 \label{eq:average_magnetization_spin_bethe}
\end{equation}
and 
\begin{equation}
 \langle m_l \rangle = \int dP_h(h_i) dP_h(h_j) 
 \left[
 \prod_{k}^{c-1} dq(u_{k\rightarrow i})  
 \prod_{m}^{c-1} dq(u_{m\rightarrow j})
 \right]
  m_l (\braces{u_{k\rightarrow i}},\braces{u_{m\rightarrow j}},h_i,h_j)
 \label{eq:average_magnetization_link_bethe}
\end{equation}

We have used the shorthand $dP_h(h_i) \equiv dh_i P_h(h_i)$ and 
$dq(u_{m\rightarrow j})\equiv du_{m\rightarrow j} q(u_{m\rightarrow j})$ to increase readability of the expressions. 


Let us focus for a moment on the magnetization functions $m_s (\braces{u_{k\rightarrow i}},h_i)$ and 
$m_l (\braces{u_{k\rightarrow i}},\braces{u_{m\rightarrow j}},h_i,h_j)$. In an actual lattice, if both
are referred to the same spin they must have the same value. This is a consequence of the consistency relations 
\eqref{eq:mi_consistency_spin}. The function $m_s$ depends on $c$ fields, of which $c-1$ are also arguments
of $m_l$. The extra field $u_{k_c\rightarrow i}$ can be obtained from the condition $m_s=m_l$. This is an 
implicit equation that we can solve. Formally this solution is written as:
\begin{equation}
 u_{k_c\rightarrow i} = \bar{u}(\braces{u_{k\rightarrow i}}^{c-1},\braces{u_{m\rightarrow j}}^{c-1},h_i,h_j)
 \label{eq:explicit_message_passing}
\end{equation}

Using the previous definition we can now relate the link magnetization to the spin one: 
\begin{equation}
 m_l(\braces{u_{k\rightarrow i}}^{c-1},\braces{u_{m\rightarrow j}}^{c-1},h_i,h_j) = 
 \int du_{k_c\rightarrow i} m_s(\braces{u_{k\rightarrow i}}^{c-1},u_{k_c\rightarrow i},h_i) \delta(u_{k_c\rightarrow i} - \bar{u}(\braces{u_{k\rightarrow i}}^{c-1},\braces{u_{m\rightarrow j}}^{c-1},h_i,h_j))
\label{eq:relation_mag_link_spin}
\end{equation}

Putting \eqref{eq:minimization_free_3}, \eqref{eq:average_magnetization_spin_bethe}, \eqref{eq:average_magnetization_link_bethe} and \eqref{eq:relation_mag_link_spin} together we get an expression that allows the determination of $q(u)$ by means of a population dynamics
scheme:

\begin{eqnarray}
\nonumber
 \lefteqn{
 \int du_{k_c\rightarrow i} dP_h(h_i) \left[\prod_{k}^{c-1} dq(u_{k\rightarrow i})\right]
 m_s (\braces{u_{k\rightarrow i}}^{c-1},u_{k_c\rightarrow i},h_i) q(u_{k_c\rightarrow i})
 =}\\&&
 \nonumber
 \int du_{k_c\rightarrow i} dP_h(h_i) dP_h(h_j) 
 \left[
 \prod_{k}^{c-1} dq(u_{k\rightarrow i})  
 \prod_{m}^{c-1} dq(u_{m\rightarrow j})
 \right]
  m_s (\braces{u_{k\rightarrow i}}^{c-1},u_{k_c\rightarrow i},h_i)\\&&
  \times
  \delta(u_{k_c\rightarrow i} - \bar{u}(\braces{u_{k\rightarrow i}}^{c-1},\braces{u_{m\rightarrow j}}^{c-1},h_i,h_j))
 \label{eq:selfconsistent_q}
\end{eqnarray}

The above equation has the following interpretation: the LHS represents the average magnetization of a spin obtained by sampling
the $c$ cavity fields $\braces{u_{k\rightarrow i}}^{c}$ from their distribution. The RHS, on the other hand, represents also the average magnetization of a spin 
but calculated by taking $c-1$ fields $\braces{u_{k\rightarrow i}}^{c}$ from their distribution and the other $u_{k_c\rightarrow i}$ fixed by the delta function to be consistent
with the link to spin marginalization. 

From \eqref{eq:selfconsistent_q} we can obtain a numerical approximation for $q(u)$ using a population 
dynamics method. The idea is to represent $q(u)$ by a list of $N\gg 1$ field values, that is, one typical sample of N values taken 
independently from $q(u)$. In order to obtain the right $q(u)$ we perform
a sampling process that simulates \eqref{eq:selfconsistent_q}. First, two groups of $(c-1)$ fields are selected randomly from the list. Consider that each set acts on one of the spins of an hypothetic link region $l$ and find the magnetization of the two spins. In this step, the disordered
external fields $h_{i}$ need to be sampled too. Once we have the magnetization predicted by the link for the spin, say $i$, we demand  
the spin region predict the same magnetization as the link. This fixes the total cavity field on the spin. Finally, substracting the $c-1$
values initially sampled from the total cavity field we get the effective cavity field $u_{k_c\rightarrow i}$ as represented in \eqref{eq:explicit_message_passing}. This value is returned to the list
of fields in a random position. The convergence of this procedure is monitored by following the evolution of the first two moments of the
list. Once the process has converged, the distribution $q(u)$ satisfying \eqref{eq:selfconsistent_q} is obtained
as a histogram of the sample list.  Algorithm (\ref{alg:BPalg}) shown below summarizes the method to follow:

\renewcommand{\algorithmiccomment}[1]{// #1}
\begin{algorithm}[H]
\caption{\textit{Population Dynamics}}
\label{alg:BPalg}
\algsetup{indent=4em}
\begin{algorithmic}[1]
\STATE Represent $q(u)$ by a list \textit{qlist} of N numbers with an arbitrary initial distribution
\STATE $i=1$
\WHILE{$i < \mbox{SWEEPMAX}$}
   \FOR[Repeat N times]{$j=1$ to N}
	\STATE Pick $(c-1)$ fields $u_{k\rightarrow i}$ randomly from \textit{qlist} and the same number of $u_{m\rightarrow j}$.
	\STATE Find $m_l$ from $m_l = Tr[\paux_i \r_l^{(ij)}]$
	\STATE From the condition $m_s=m_l$ obtain a new field $u_{k_c\rightarrow i}$
	\STATE Put the new $u_{k_c\rightarrow i}$ back to \textit{qlist}, substituting one element chosen at random	
   \ENDFOR
   \STATE Check the first and second moments of \textit{qlist}
   \IF{Relative change of moments is smaller
   than $\mbox{TOL} = 10^{-4}$}
	\STATE Calculate observables $O[q(u)]$ by sampling \textit{qlist} repeatedly
	\STATE $i = \mbox{SWEEPMAX}$ \COMMENT {\hspace{1cm}Stop iterations}
   \ENDIF
   \STATE i++
\ENDWHILE
\RETURN
\end{algorithmic}
\end{algorithm}

%

\subsection{Kikuchi approximation}

The formalism for the plaquette approximation is essentially the same. We will assume for definiteness a 2D
configuration but the results are easy to extend to more dimensions or other kind of lattices, for example, triangular ones.
The intensive variational free energy according to \eqref{eq:average_free_cvm_3} is:
\begin{equation}
 -\beta f_{\text{Kik}} = \langle \ln \partition_p \rangle -2 \langle \ln \partition_l \rangle + \langle \ln \partition_s \rangle
 \label{eq:variational_free_energy_kikuchi}
\end{equation}

Below we include for clarity the resulting equations for all the $\langle \ln\mathcal{Z}_{r}\rangle$ for a spin region $s=(i)$,
a link $l=(i,j)$ and a plaquette $p = (i,j,k,m)$:

\begin{equation}
 \label{eq:average_partition_spin_kikuchi}
 \langle \ln\mathcal{Z}_{s}\rangle = 
 \int 
     dP_h(h_i) 
     \left[\prod_{l'\in\mathcal{P}(i)} dq(u_{l'\rightarrow i})\right] 
     \ln Tr\left[ \exp -\beta\left(\ham_s^{\text{eff}}(\pau_i) \right) \right]
\end{equation}

\begin{eqnarray}
\nonumber
 \langle \ln\mathcal{Z}_{l}\rangle &=& 
 \int 
    \prod_{i'\in \mathcal{D}(l)}
    \left[dP_h(h_{i'})\prod_{l' \in \mathcal{P}(i')\setminus l}  dq(u_{l'\rightarrow i'}) \right]
    \left[\prod_{p' \in \mathcal{P}(l)}  dQ(U_{p'\rightarrow l},u_{p'\rightarrow i}, u_{p'\rightarrow j})\right] \\
 &\times& \ln Tr\left[\exp -\beta
      \left(       
      \ham_l^{\text{eff}}(\pau_{i},\pau_{j})
      \right)\right]
      \label{eq:average_partition_link_kikuchi}
\end{eqnarray}

\begin{eqnarray}
\nonumber
 \langle \ln\mathcal{Z}_{p}\rangle &=& 
 \int 
    \prod_{i'\in \mathcal{D}(p)}\left[ dP_h(h_{i'}) \prod_{\substack{l' \in \mathcal{P}(i')\\ l' \notin \mathcal{D}(p)}}  dq(u_{l'\rightarrow i'}) \right]
    \left[\prod_{\substack{ l\in \mathcal{D}(p)\\ l = (i_l, j_l) }}\prod_{p' \in \mathcal{P}(l)\setminus p}  dQ(U_{p'\rightarrow l},u_{p'\rightarrow i_{l}}, u_{p'\rightarrow j_{l}})\right] \\
 &\times& \ln Tr\left[\exp -\beta
      \left(       
      \ham_p^{\text{eff}}(\pau_{i},\pau_{j},\pau_{k},\pau_{m})
      \right)\right]
      \label{eq:average_partition_plaq_kikuchi}
\end{eqnarray}

To lighten the formulas above we have used again the convention $dq(u_{l\rightarrow i}) \equiv du_{l\rightarrow i}  q(u_{l\rightarrow i})$ and $dQ(U_{p\rightarrow l},u_{p\rightarrow i}, u_{p\rightarrow j}) \equiv du_{p\rightarrow i} du_{p\rightarrow j} dU_{p\rightarrow l} Q(U_{p\rightarrow l},u_{p\rightarrow i}, u_{p\rightarrow j})$. Notice that
the field probability distribution of each region is factored in terms of single $q(u_{l\rightarrow i})$ and $Q(U_{p\rightarrow l},u_{p\rightarrow i}, u_{p\rightarrow j})$. The real interactions are put together  with the cavity ones into an effective hamiltonian that includes all the terms of the Bethe case plus the plaquette-to-link fields:

\begin{eqnarray*}
 \ham_s^{\text{eff}}(\pau_i)&=& \ham_s(\pauz_i) - \paux_{i} \sum_{l'\in\mathcal{P}(i)}   u_{l'\rightarrow i}\\
 \ham_l^{\text{eff}}(\pau_{i},\pau_{j}) &=& 
       \ham_l(\pau_{i},\pau_{j}) 
      -\sum_{i'\in \mathcal{D}(l)} \paux_{i'} \sum_{l' \in \mathcal{P}(i')\setminus l} u_{l'\rightarrow i'}
      -\sum_{p' \in \mathcal{P}(l)} [U_{p'\rightarrow l} \paux_{i} \paux_{j} + u_{p'\rightarrow i} \paux_{i} + u_{p'\rightarrow j} \paux_{j}]\\
 \ham_p^{\text{eff}}(\pau_{i},\pau_{j},\pau_{k},\pau_{m}) &=& 
       \ham_p(\pau_{i},\pau_{j},\pau_{k},\pau_{m}) 
      -\sum_{i'\in \mathcal{D}(p)} \paux_{i'} \sum_{\substack{l' \in \mathcal{P}(i')\\ l' \notin \mathcal{D}(p)}} u_{l'\rightarrow i'}
      -\sum_{\substack{ l\in \mathcal{D}(p)\\ l = (i_l, j_l) }}\sum_{p' \in \mathcal{P}(l)\setminus p} [U_{p'\rightarrow l} \paux_{i_l} \paux_{j_l} + u_{p'\rightarrow i_l} \paux_{i_l} + u_{p'\rightarrow j_l} \paux_{j_l}]
\end{eqnarray*}

Now we use the stationarity conditions \eqref{eq:stationarity_condition_kikuchi} to obtain the relation between the first and second average 
moments (\ie the magnetization and correlation) predicted by each region:

\begin{eqnarray}
 \label{eq:minimization_free_1_kikuchi}
 \langle m_s \rangle &=& \langle m_l\rangle = \langle m_p\rangle\\
 \langle c_l \rangle &=& \langle c_p\rangle
 \label{eq:minimization_free_2_kikuchi}
\end{eqnarray}

From the first equality in \eqref{eq:minimization_free_1_kikuchi} and repeating the steps for the Bethe case we get
an expression for the $q$ distribution:

\begin{eqnarray}
\nonumber
 \lefteqn{
 \int du_{k_c\rightarrow i} dP_h(h_i) \left[\prod_{k}^{c-1} dq(u_{k\rightarrow i})\right]
 m_s (\braces{u_{k\rightarrow i}}^{c-1},u_{k_c\rightarrow i},h_i) q(u_{k_c\rightarrow i})
 =}\\&&
 \nonumber
 \int du_{k_c\rightarrow i} dP_h(h_i) dP_h(h_j) 
 \left[
 \prod_{k}^{c-1} dq(u_{k\rightarrow i})  
 \prod_{m}^{c-1} dq(u_{m\rightarrow j})
 \prod_{p \in \mathcal{P}(l)}^{2}  dQ(U_{p\rightarrow l},u_{p\rightarrow i}, u_{p\rightarrow j})
 \right]
  m_s (\braces{u_{k\rightarrow i}}^{c-1},u_{k_c\rightarrow i},h_i)\\&&
  \times
  \delta\left[u_{k_c\rightarrow i} - \bar{u}\left(\braces{u_{k\rightarrow i}}^{c-1},\braces{u_{m\rightarrow j}}^{c-1}
                                               ,\braces{U_{p\rightarrow l},u_{p\rightarrow i}, u_{p\rightarrow j}}^2,h_i,h_j\right)\right]
 \label{eq:selfconsistent_q_kikuchi}
\end{eqnarray}

In \eqref{eq:selfconsistent_q_kikuchi} the function $\bar{u}$ is the one which gives the effective field that makes
the magnetization predicted by the spin consistent with the magnetization predicted by the link. Compared to the Bethe case it
now includes the dependence on the plaquette-to-link messages $U_{p\rightarrow l},u_{p\rightarrow i}, u_{p\rightarrow j}$.

Let us work now with the rightmost equation of \eqref{eq:minimization_free_1_kikuchi} or, equivalently, with \eqref{eq:minimization_free_2_kikuchi}.
These involve the consistency between plaquettes and links. It is to be noted that
the single-instance update equations based on a plaquette to link marginalization \textit{alone} do not determine the value of the fields $u_{p\rightarrow i}$.
Instead, only the sum $u_{p\rightarrow i} + u_{l\rightarrow i}$ and $u_{p\rightarrow j} + u_{l\rightarrow j}$ is completely specified. As a a consequence, equations 
\eqref{eq:minimization_free_1_kikuchi} or \eqref{eq:minimization_free_2_kikuchi} will not give us an expression for 
$Q(U_{p\rightarrow l},u_{p\rightarrow i}, u_{p\rightarrow j})$. Alternatively, we get an equation for the joint distribution $R(U_{p\rightarrow l},u_i, u_j)$ of the correlation field $U_{p\rightarrow l}$ and the sum of the magnetization fields $u_i=u_{p\rightarrow i} + u_{l\rightarrow i}$ and $u_j=u_{p\rightarrow j} + u_{l\rightarrow j}$. This distribution is defined as the convolution of the original $Q$ and $q$,
\begin{equation}
 R(U_{p\rightarrow l},u_i, u_j) = \int du_{p\rightarrow i} du_{p\rightarrow j} 
 Q(U_{p\rightarrow l},u_{p\rightarrow i}, u_{p\rightarrow j})
 q(u_i - u_{p\rightarrow i} ) q(u_j - u_{p\rightarrow j})
\end{equation}
and obeys an equation that is structurally very similar to \eqref{eq:selfconsistent_q_kikuchi}:

\begin{eqnarray}
 \nonumber
 \langle c_l\rangle&=&\langle c_p\rangle\\
 \nonumber
 \langle c_l\rangle&=&  
 \int 
 dU_{p\rightarrow l}du_{i}du_{j}
 dP_h(h_i) dP_h(h_j) dR(U_{p'\rightarrow l},u'_i, u'_j) dq(u_{l'_1\rightarrow i}) dq(u_{l'_2\rightarrow j})\\
 \nonumber
 &\times&c_l(\#_l,U_{p\rightarrow l},u_{i},u_{j}) R(U_{p\rightarrow l},u_i, u_j)\\
\nonumber
 \langle c_p\rangle&=& 
 \int 
    dU_{p\rightarrow l}du_{i}du_{j}
    \prod_{i'\in \mathcal{D}(p)}\left[ dP(h_{i'}) \prod_{\substack{l' \in \mathcal{P}(i')\\ l' \notin \mathcal{D}(p)}}  dq(u_{l'\rightarrow i'}) \right]
    \left[\prod_{l''\in \mathcal{D}(p)}\prod_{p' \in \mathcal{P}(l'')\setminus p}  dQ(U_{p'\rightarrow l''},u_{p'\rightarrow i_{l''}}, u_{p'\rightarrow j_{l''}})\right] \\    
 \nonumber
 &\times& c_l(\#_l,U_{p\rightarrow l},u_{i},u_{j})  \\
 &\times& \delta\left[u_{i} - \bar{u}_i\left(\#_p\right)\right]
	  \;\;
          \delta\left[u_{j} - \bar{u}_j\left(\#_p\right)\right]
          \;\;
          \delta\left[U_{p\rightarrow l} - \bar{U}_{p\rightarrow l}\left(\#_p\right)\right]                                            
 \label{eq:selfconsistent_R_kikuchi}
\end{eqnarray}

In the above equation, the symbol $\#_p$ in the RHS stands for all the cavity fields acting on the plaquette $p$ from neighboring regions. It includes
also the local magnetic field on each spin. The symbol $\#_l$ includes a subset of $\#_p$; just those fields acting on the link $l\in p$.
The functions $\bar{U}_{p\rightarrow l}\left(\#_p\right)$, $\bar{u}_i\left(\#_p\right)$ and $\bar{u}_j\left(\#_p\right)$ give the effective correlation and magnetization fields on link $l$ due to the interactions in plaquette $p$.
\footnote{The expressions $\bar{U}$, $\bar{u}_i$ and $\bar{u}_j$ 
are again only formal representations of the result of the self-consistent determination of the fields.}

\section{Numerical Results}

\subsection{Quantum Transverse Ising Model}

For an homogeneous system $J_{ij}=J$, in an homogeneous transverse field $h_i=h$, the numeric solution of the update equations in single instances simplifies significantly. The model at equilibrium should be described  by a fixed point iteration of a single combination of the parameters $u_{l\rightarrow i}$, $u_{p\rightarrow l}$ and $U_{p\rightarrow l}$. Therefore we can take
only one pair of plaquette-to-link and link-to-spin marginalization equations and iterate them recursively until the $(u_{l\rightarrow i},u_{p \rightarrow l},U_{p \rightarrow l})$ combination reaches a fixed point. To simplify the notation we also drop specific spatial indexes and write only $(u_{l},u_{p},U_{p})$.

Our description below makes emphasis on the plaquette approximation considering that the Bethe case is extensively presented in a vast literature. All the same, the reader interested only on the Bethe approximation can formaly put $U_{p}$ and $u_{p}$ to zero and iterate only the link-to-spin marginalization condition.

The sequence of steps is the following. First, the $(u_{l},u_{p},U_{p})$ triad is initialized
to some arbitrary real values. Then these fields are used to evaluate the moments of $\r_p^{(ijkm)}$ of an imaginary plaquette: 
\begin{eqnarray}
 m_p &=& Tr[\paux_i \r_p^{(ijkm)}]  \label{eq:mi_consistency_2} \\
 c_p &=& Tr[\paux_i \paux_j \r_p^{(ijkm)}]  \label{eq:corr_consistency_2} 
\end{eqnarray}
Calculations are made only for  one spin and one link for symmetry reasons. As a consequence of \eqref{eq:mi_consistency_spin} and \eqref{eq:mi_consistency} a new value for the field $u'_{l}$ is generated such that spin magnetization $m_i = Tr[\paux_i \r_s^{(i)}]$ equals $m_p$. Explicitly, the equation that must be solved is:
$$m_p = \dfrac{Ku'_{l}}{\sqrt{h^2 + {(Ku'_{l})}^2}}\tanh \beta \sqrt{h^2 + {(Ku'_{l})}^2}$$
where $K=4$ is the connectivity of the spin in 2D. After, new values  $u'_{p}$ and $U'_{p}$ are obtained from the LHS of \eqref{eq:mi_consistency} and \eqref{eq:corr_consistency}. Their value must make the magnetization and correlation from $\r_l^{(ij)}$ consistent with \eqref{eq:mi_consistency_2} and \eqref{eq:corr_consistency_2}.
We have:
\begin{eqnarray}
 m_l &\equiv& Tr[\paux_i \r_l^{(ij)}] = f(u'_{l},u'_{p},U'_{p}) \label{eq:mi_consistency_3} \\
 c_l &\equiv& Tr[\paux_i \paux_j \r_l^{(ij)}] = g(u'_{l},u'_{p},U'_{p})  \label{eq:corr_consistency_3} 
\end{eqnarray}
and then we would have to solve:
\begin{eqnarray}
f(u'_{l},u'_{p},U'_{p}) &=& m_p \label{eq:mi_consistency_4} \\
g(u'_{l},u'_{p},U'_{p}) &=& c_p \label{eq:corr_consistency_4} 
\end{eqnarray}

In the equations above the value of $u'_{l}$ from the previous step is used when solving for $u'_{p},U'_{p}$. This algorithm is repeated until
stability is reached \ie until the variation of the field values drop below certain prefixed threshold. For each temperature and/or field it is convenient to use as initial values the results obtained for a nearby point in the phase diagram. This improves the convergence speed significantly.

To compare this result with actual message passing equations in single instances we studied a 16x16 square lattice with homogeneous field and periodic boundary conditions iterating \eqref{eq:mi_consistency_spin}-\eqref{eq:corr_consistency} starting from random initial conditions and following a random update scheme until convergence (SI). 
Since the system is homogeneous, there is no need of running a large number of instances nor using a large system; field values tend
to the same value everywhere. The only difference between samples would be the initial conditions and the actual random update order. In this case we averaged results for 10 different initial field configurations. 

Although   the population dynamics (PD) solution to the problem is introduced properly in the next section, since it is mainly relevant for disordered systems, its application to this model is shown here for completeness.  Broadly speaking, it is similar to the FP but focuses on the stability of a population of cavity fields instead of a single set of values. The population of field values is supposed to represent the distribution of fields for the average case scenario. In the homogeneous system populations will be represented by a single value, this is, distributions are delta-shaped around the fixed point fields.

\begin{figure}[!htb]
\centering
\subfloat[]{
\includegraphics[width=0.45 \textwidth,keepaspectratio=true]{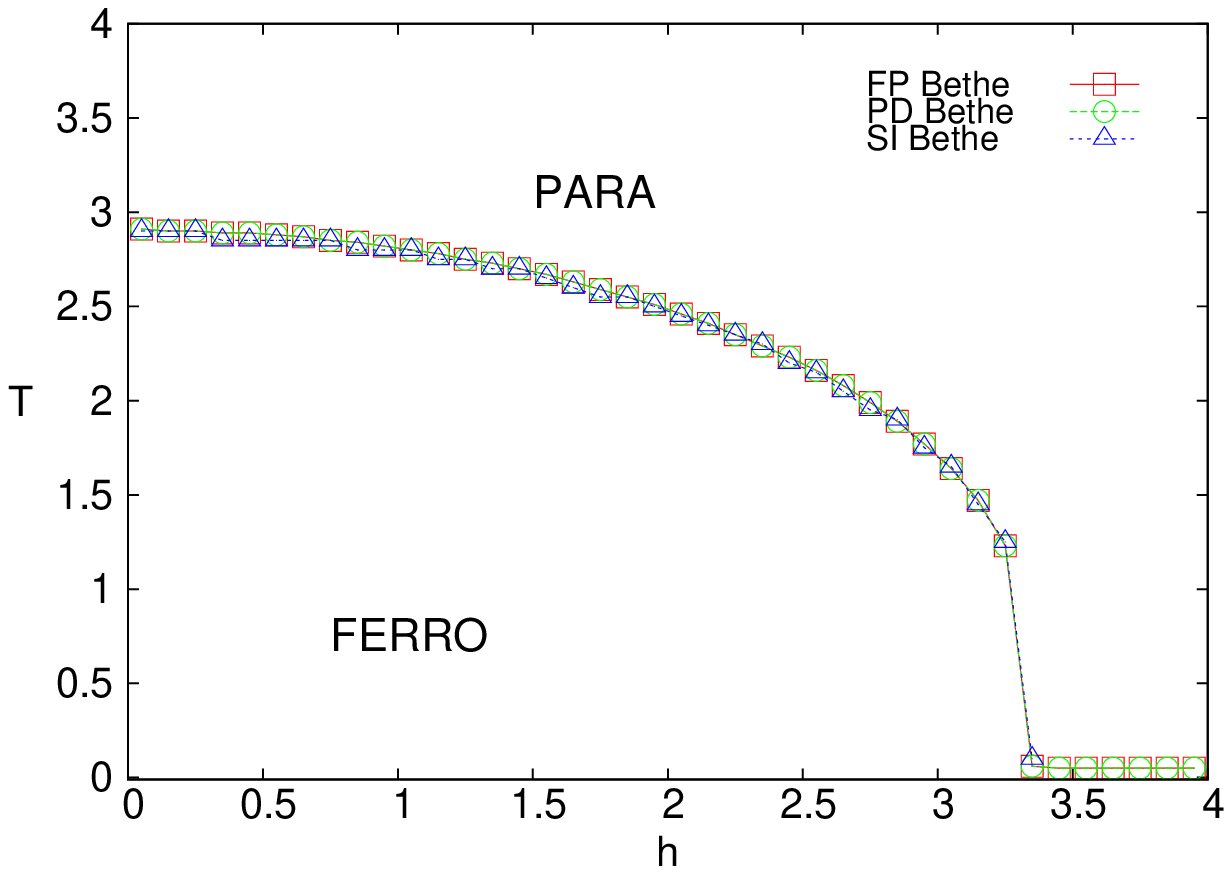}}
\subfloat[]{
\includegraphics[width=0.45 \textwidth,keepaspectratio=true]{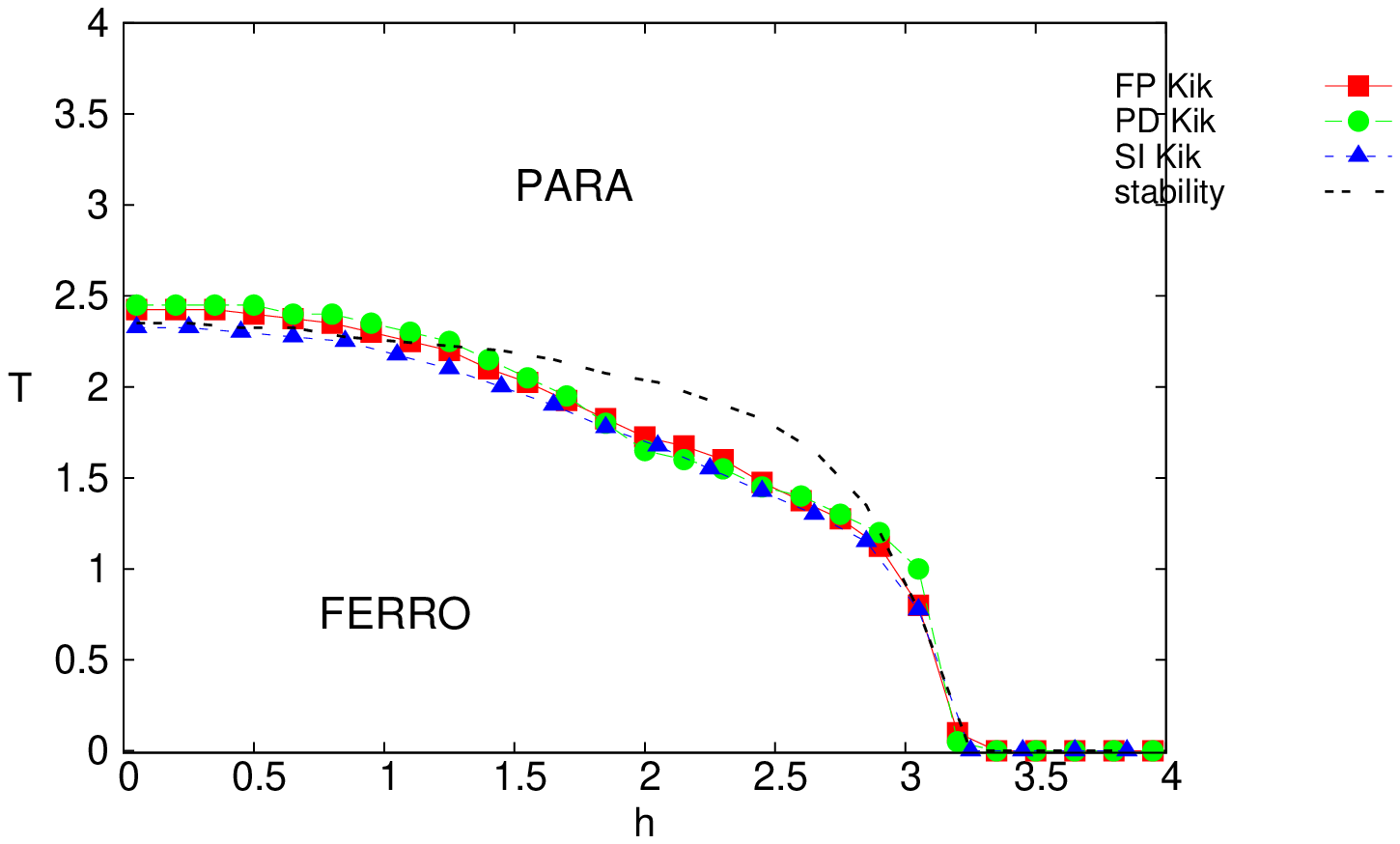}}

\caption{ $h-T$ phase diagram of the transverse field Ising model in two dimensions. Three methods are considered: simulations in single instances (SI), fixed point iterations (FP) and population dynamics (PD) (See text for details). (a)Bethe approximation (b)Plaquette approximation. 
Both approximations reproduce the $h\rightarrow 0$ classical limit and find an estimate of the $h_c$ above which the ordered phase disappear.}
\label{fig:ferro_bethe_kikuchi_pd}
\end{figure}

The numerical results obtained for the three methods are shown in Fig.(\ref{fig:ferro_bethe_kikuchi_pd}). In this figure we present the H-T phase diagram of the Quantum Ising model in a transvere magnetic field for the Bethe and Kikuchi approximations.

From Fig.(\ref{fig:ferro_bethe_kikuchi_pd}) we observe that in all the  approximations we get a line dividing a paramagnetic solution where
the spontaneous magnetization in the $\hat x$ direction is zero from another region where long range order dominates.
For low transverse field the transition temperature coincides with the classical case prediction. We also find that above a critical value of
the external field quantum fluctuations destroy the possibility of ferromagnetic order at any temperature. The results for the Bethe case are a lot less noisy that for the plaquette approximation. It is interesting that in the latter all methods find a region for intermediate values of the external field where there is a gap of non-convergence between the paramagnetic region an the ferromagnetic one. It is not clear to us whether it is a numerical problem or if it is an intrinsic property of the approximation. 

\begin{figure}[h]
 \centering
 \subfloat[Bethe approximation]{
 \includegraphics[width=0.45 \textwidth,keepaspectratio=true]{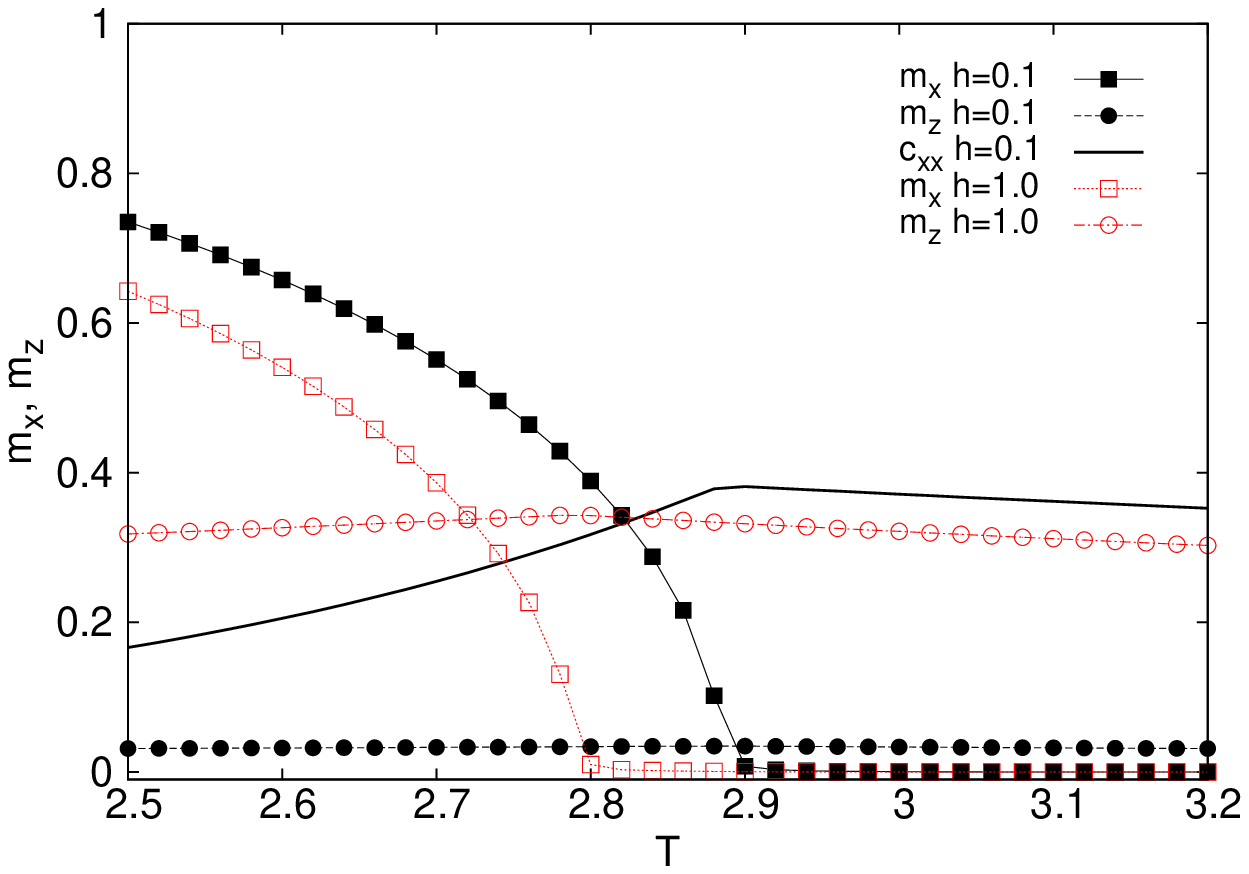}}
 \subfloat[Kikuchi approximation]{
ca \includegraphics[width=0.45 \textwidth,keepaspectratio=true]{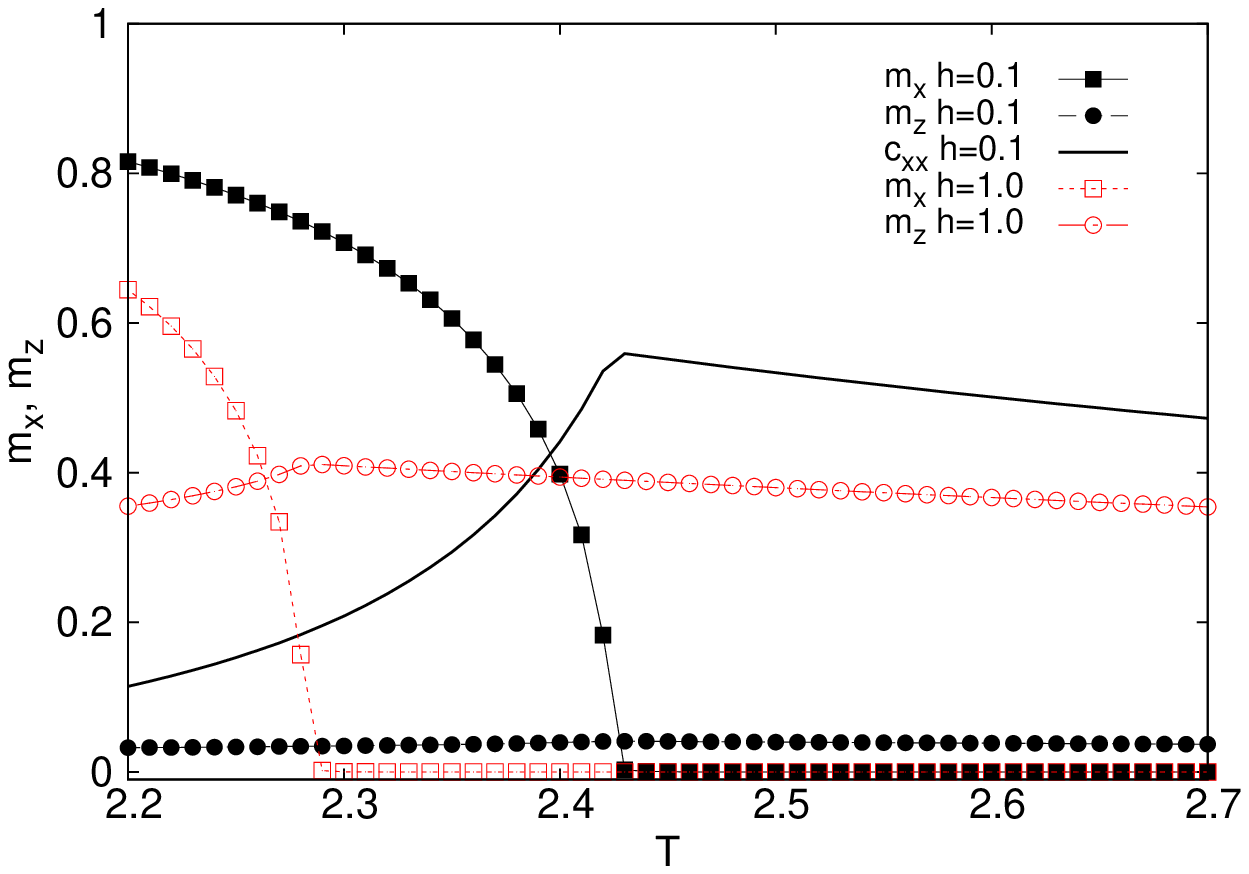}}
 \caption{Temperature dependence of the magnetization and correlation for two representative values of the external field in the a) Bethe and b) Kikuchi approximation.
 We present the transverse value ($m_z$, circles) and the longitudinal one($m_x$, squares). It is in the longitudinal axis where long range order emerges by lowering the temperature for a fixed field.  A low field value (full symbols) does not affect strongly the transition temperature observed for the $m_x$ magnetization at $h=0$. This corresponds to the Bethe critical temperature
 of the classical model, $T_c\approx2.89$ in part a) and to $T_c\approx 2.43$ in the Kikuchi approximation in part b). Higher transverse fields make the transition move towards lower temperatures under the influence of quantum fluctuations. The transverse magnetization $m_z$ reflects the phase transition in the form of a small kink. The connected correlation in the $\hat x$ direction,  $c_{xx}=<\paux_i \paux_j> - <\paux_i> <\paux_j>$, has a maximun in the vicinity of the transition in both cases.}
 \label{fig:mxmzbethe}
\end{figure}

In Fig.(\ref{fig:mxmzbethe}) we show vertical cuts of the phase diagram taken at $h=0.5$ and $h=2.5$. The behavior of the longitudinal magnetization is qualitatively equivalent to the classical ferromagnetic case. In the transversal direction the system
presents always a magnetization in the same direction of the applied field. In these plot we see again that for small fields the 
critical temperature depends weakly on $h$, taking values very close to the classical one. As the external fields increases, the system needs to lower the temperature to establish the long range order. This can be done up to a certain critical field, above which
quantum fluctuations destroy the possibility of an ordered phase.

\subsection{Quantum Transverse Ising model in a Random  Field}

To numerically approach the average case scenario at the Kikuchi level is not as simple as in the Bethe case a result that is already known from classical models\cite{tommaso_CVM}. The problem relies in the proper equations \eqref{eq:selfconsistent_q_kikuchi} and \eqref{eq:selfconsistent_R_kikuchi} that do not define a simple closed equation for $Q$ and to obtain it obtaining by deconvolving $R$ and $q$ is numerically challenging. Further complications may also arise from the fact that $Q$ need not to be positive definite. Nonetheless, one can still study the average properties of the update equations in the Kikuchi approximation using population dynamics. The result, though, will be a solution of  \eqref{eq:selfconsistent_q_kikuchi} and \eqref{eq:selfconsistent_R_kikuchi} only in the paramagnetic regime. In the case of distributions with permanent magnetization, it constitutes
only an heuristic tool.

In a 2D square lattice, a plaquette $p$ has four other neighbor plaquettes $p'$ with which it shares a link. From each of these regions it interacts
via one plaquette to link triplet $(U_{p'\rightarrow l},u_{p'\rightarrow i},u_{p'\rightarrow j})$ and two link-to-spin fields $u_{l'\rightarrow i}$,
$u_{l''\rightarrow j}$, five fields in total. Given all those external messages one can, using the update equations, find the fields inside the plaquette. In order to keep as much information as possible we define a population representing the joint distribution of the five values mentioned before. Following the scheme for the Bethe case, we sample the surroundings of the plaquette, calculate a new set of messages and return it to the population. Once the population stabilizes, all the observables can be found by sampling repeatedly the resulting distribution.

\begin{figure}[h]
 \centering
 \includegraphics[width=0.5\textwidth,keepaspectratio=true]{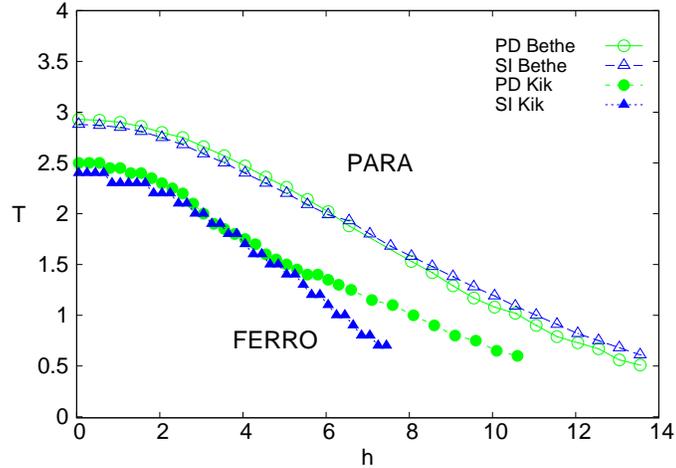}
 \caption{Quantum CVM for transverse RFIM. $h-T$ phase diagram of the RFIM model in two dimensions. The disordered field
 on each site is taken randomly in the $[0,h)$ interval. Two methods are used: simulations in single instances (SI) and population dynamics (PD) (See text for details). In the Kikuchi approximation convergence is a serious issue near the critical line. In fact what is shown in the figure for the plaquette case is the line where the ferromagnetic solution looses stability and the algorithm
 stops converging. The convergence is recovered later inside the paramagnetic phase.}
 \label{fig:rfim_results}
\end{figure}

\begin{figure}[h]
 \centering
 \subfloat[Distribution of $u_{l\rightarrow i}$ fields. Bethe approximation]{
    \includegraphics[width=0.45\textwidth,keepaspectratio=true]{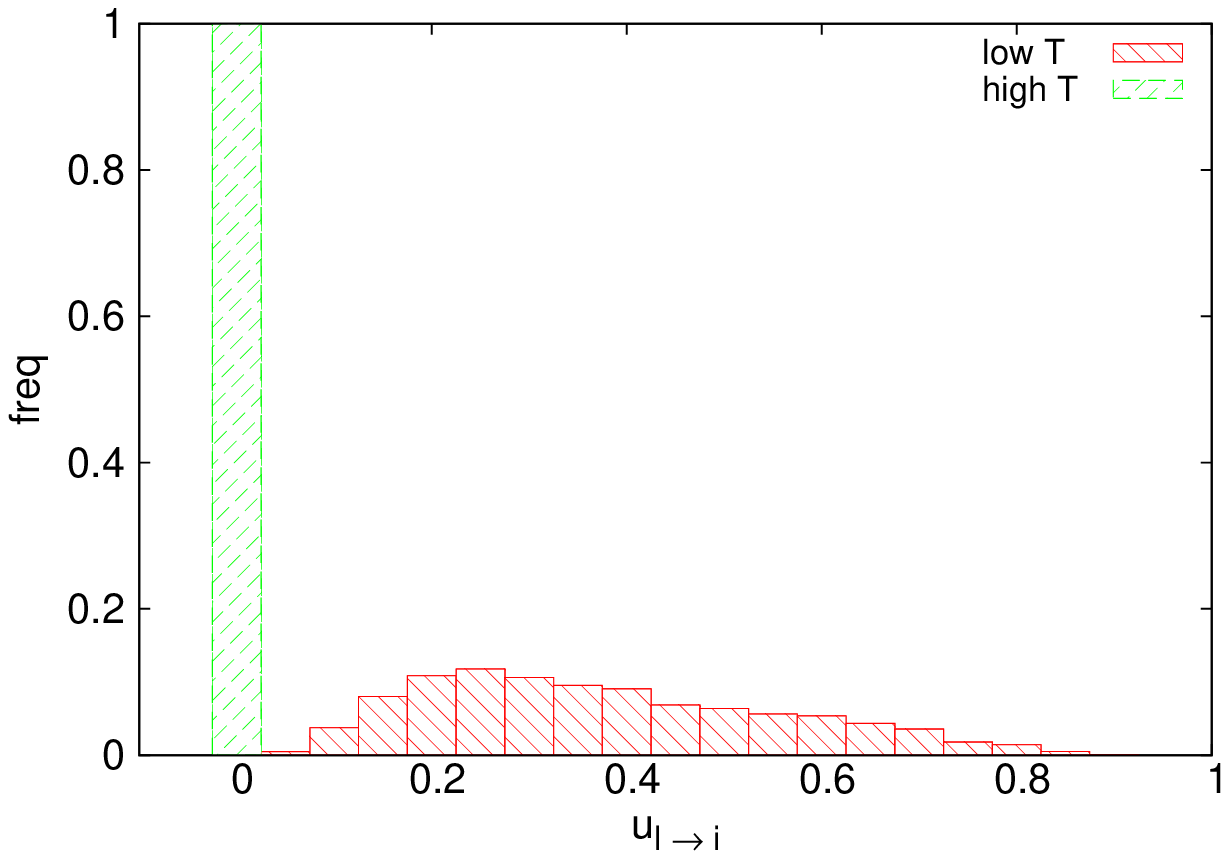}
    \label{fig:rfim_uLdist}
 }
 \subfloat[Distribution of $U_{p\rightarrow l}$ fields. Kikuchi approximation]{
    \includegraphics[width=0.45\textwidth,keepaspectratio=true]{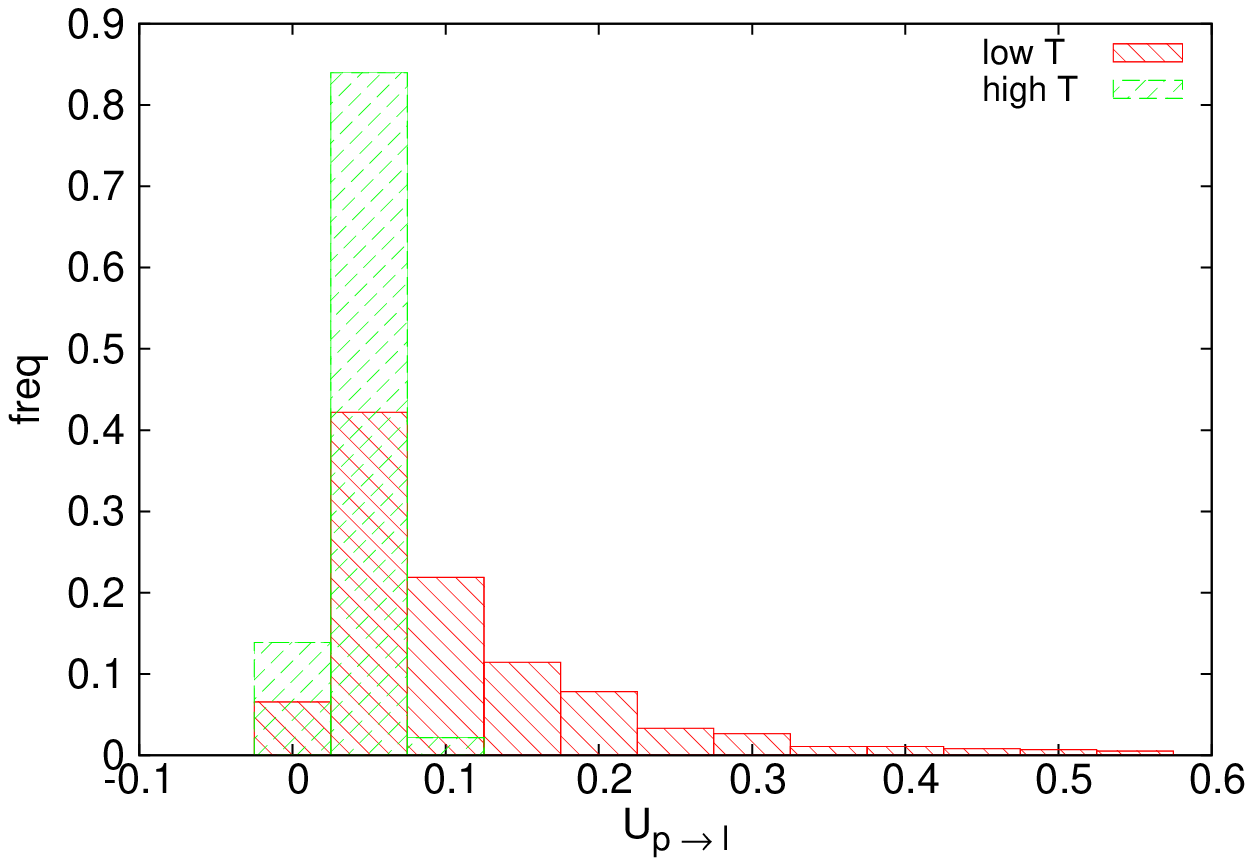}
    \label{fig:rfim_Udist}
}
 \caption{
    These histograms show some of the relevant effective field distributions for the Bethe approximation (left) and Kikuchi's (right) at two different temperatures for a given field intensity. The high temperature distribution corresponds to the point ($h=4.0, T=3.5$), well in the paramagnetic phase of Fig.(\ref{fig:rfim_results}). On the other hand, the low temperature data was obtained
    for the same field at $T=1.0$, inside the ferromagnetic phase in both approximations. For the magnetization fields $u_{l\rightarrow i}$ in a) we observe that at high temperature the distribution is in fact a delta function around zero that spreads when moving into lower temperature regions.
    The $U_{p\rightarrow l}$ distribution for the plaquette approximation in b) peaks around a non-zero value for high temperatures, when the system is
    spatially homogeneous in the $\hat x$ direction. In the paramagnetic region this distribution spreads after the onset of heterogeneous local magnetizations.
}
\end{figure}

In Fig.(\ref{fig:rfim_results}) we compare the results of using the population dynamics algorithm and single instance simulations for the RFIM.
Similar to the ordered case, the phase diagram of the $\hat x$ magnetization is divided in two regions, para and ferromagnetic. The classical 
limits of low fields are in agreement with the previously known results and of course with the corresponding values in Fig.(\ref{fig:ferro_bethe_kikuchi_pd}). For the SI calculations, 100 samples of a 32x32 square lattice  with periodic boundary conditions are averaged.
For high $h$ values convergence is an issue for both PD and SI simulations. Also, in this region the longitudinal $m_x$ magnetization is rather
small in the ferromagnetic region. We did not managed to observe a critical $h_c$ value as in the ordered model.

The shape of the field distributions on the lattice changes for the para or ferromagnetic phase. In the paramagnetic region the magnetization fields,
$u_{l\rightarrow i}$ and $u_{p\rightarrow i}$ distribute as delta functions around zero, see for example Fig.(\ref{fig:rfim_uLdist}). The correlation
fields $U_{p \rightarrow l}$ are also well centered around a given value for high temperatures, see Fig.(\ref{fig:rfim_Udist}). On the other hand,
inside the ferromagnetic phase, due to the heterogeneous local fields in the $\hat z$ direction, we observe that all distributions spread suggesting the possible existence of a glassy phase.

\section{Conclusions}

In this work, we first re-derived the equations for the Cluster Variational Method for models involving quantum phase transitions. Starting from a variational expression for a region based free energy we managed to find approximations to local probability distributions. The minimization of the region free energy is somewhat hindered by the quantum nature of the hamiltonian and the non-conmutativity of the operators appearing on it. As a consequence the cavity fields of the classical models transform in our approach into hermitian operators, parametrized by Pauli matrices. We then approximate the problem transforming these equations for operators into an approximate set of equations for the parameters describing the density operators defining the variational method.

This quantum-CVM is a good framework for studying finite dimensional models. For ordered systems the standard approach exploits the translational symmetry and reduce the problem to the determination of a handful of parameters a technique very well known in the literature. On the other hand, for disordered models we were able to transform the consistency relations imposed between overlapping regions into proper message passing equations that can be treated in polinomial time. We showed by studying the Quantum Ising model in a transverse uniform external field, that both approaches are equivalent when disorder is absent. When disorder is present, like in the Quantum Ising model in a transverse random external field, the message passing equations derived here become nevertheless a very efficient computational approach to study the properties of the model.

In a more general setting, in this work we also presented  a version of the CVM for quantum models within an average case scenario, i.e. where the average over the disorder is done without specifically treating single instances.  Although this generalization translates into a very complex set of population dynamic equations between operators, we can approximate them through complex populations of physically sound parameters, here magnetization and correlations that can be solved using a variation of standard techniques. The results of all the approaches were compared studying the Quantum Ising model in a transverse random external field.

\bibliography{bibqcvm}

\end{document}